\documentclass[%
preprint,
superscriptaddress,
 amsmath,amssymb,
 aps,
]{revtex4-1}
\usepackage{graphicx}
\usepackage{dcolumn}
\usepackage{bm}
\usepackage{subfigure}
\usepackage{mathrsfs} 
\usepackage{epsfig}
\usepackage{epstopdf}
\usepackage{color}

\newcommand{\abs}[1]{\left|{#1}\right|}

\newcommand{\eq}[1]{Eq.~(\ref{#1})}
\newcommand{\fig}[1]{Fig.~\ref{#1}}

\def\beq{\begin{eqnarray}}
\def\eeq{\end{eqnarray}}

\newcommand{\expec}[1]{\langle#1\rangle}

\newcommand{\re}[1]{\textrm{Re}[#1]}
\newcommand{\im}[1]{\textrm{Im}[#1]}

\begin{document}

\title{Robust squeezed light against mode mismatch using a self imaging optical parametric oscillator}

\author{Chan Roh}
\author{Geunhee Gwak}
\author{Young-Sik Ra}
 \email{youngsikra@gmail.com}
\affiliation{Department of Physics, Korea Advanced Institute of Science and Technology, Daejeon 34141, Korea}%
\date{\today}

\begin{abstract}
We present squeezed light that is robust against spatial mode mismatch (beam displacement, tilt, and beam-size difference), which is generated from a self-imaging optical parametric oscillator below the threshold. We investigate the quantum properties of the generated light when the oscillator is detuned from the ideal self-imaging condition for stable operation. We find that the generated light is more robust to mode mismatch than single-mode squeezed light having the same squeezing level, and it even outperforms the single-mode infinitely squeezed light as the strength of mode mismatch increases.
\end{abstract}

\flushbottom
\maketitle

\thispagestyle{empty}

\section{Introduction}

Squeezed light is a versatile quantum resource for quantum information technologies~\cite{Asavanant:2019iw,Larsen:2019gg,Ra:2020gg,Pfister:2019ck,Zhong:2020cu,Arrazola:2021fx,Liu:2020gh}. In particular, in quantum metrology, it reduces noises in measurement below the standard quantum limit~\cite{giovannetti04,taylor13,Guo:2019el}; a notable application is gravitational wave detectors~\cite{Tse:2019jp,Acernese:2019hg}. The quality of squeezed light is typically measured by the squeezing level (the degree of noise reduction compared with the vacuum noise), which is reported up to 15 dB by using an optical parametric oscillator (OPO)~\cite{Vahlbruch:2016jf}.

To exploit the full potential of squeezed light in quantum technologies, however, a precise mode matching with subsequent operations is required~\cite{Oelker:2014kv,Toyra:2017bg,steinlechner18}. Mismatch of modes results in a loss of original properties of light, e.g. degradation of the squeezing level~\cite{Toyra:2017bg}. For classical light, mode mismatch is not a critical issue because the loss of light can be simply compensated by means of increasing the optical power. However, for squeezed light, there is a limit on compensation by increasing the initial squeezing level, e.g., if the mode matching efficiency is less than 50 \%, there is no way to obtain more than 3 dB squeezing. Moreover, if mode mismatch varies dynamically or even fluctuates stochastically, its correction is a very challenging problem~\cite{Oelker:2014kv,Wittel:2015wy}.

A way of circumventing this problem is to prepare squeezed light in a few additional modes, so that, in case of mode mismatch, squeezed light from the additional modes is coupled into a target mode~\cite{lugiato97,embrey15}. Recently, such an idea is demonstrated by employing an additional OPO operating in a Hermite-Gaussian HG$_{01}$  mode for a vertical mode mismatch~\cite{steinlechner18}. However, to be robust against more general and stronger mode mismatch, it is required to build many phase-locked OPOs, each operating in a different HG$_{mn}$ mode~\cite{Hsu:2004kz}, which is technically demanding.

Here we show that a single OPO in a self-imaging configuration can generate squeezed light that is robust against various cases of mode mismatch. The self-imaging OPO is based on a fully degenerate optical cavity in spatial modes, called a self-imaging cavity~\cite{Arnaud:1969gh,chalopin10}, hence it can support spatially multimode quantum states generated by a parametric down-conversion process~\cite{lopez09,Chalopin:2010co,chalopin11}. However, the ideal self-imaging condition leads to cavity instability, and thus, a small detuning is necessary for stable operation~\cite{chalopin10}. In this work, we characterize the quantum properties of the light from a self-imaging OPO with a small detuning (parametrized by the Gouy phase shift) and an intracavity loss, and we find that squeezed vacua at multiple HG$_{mn}$ modes are simultaneously generated  without much degradation from the ideal self-imaging condition. We then investigate the injection of the generated multimode light into a target mode in the presence of mode mismatch (beam displacement, tilt, and beam-size difference both for horizontal and vertical directions) and optical losses. We find that the multimode squeezed light is more robust to mode mismatch than single-mode squeezed light with the same squeezing level, and at sufficiently large mode mismatch, it even outperforms the single-mode infinitely squeezed light. Such robustness to mode mismatch can be achieved in a detuning range of a self-imaging OPO within reach of the experimental control.


\section{Quantum properties of multimode squeezed light} \label{section:multisqz}

\subsection{Spatial properties of parametric down conversion}

To generate squeezed light, we consider a degenerate type-I parametric down conversion in a $\chi^{(2)}$nonlinear crystal in the low gain regime~\cite{Caspani:2010jq}. The interaction Hamiltonian can be described as
\begin{equation}
	\hat{H} = -{i\hbar{g} \over 2} \int d^2\vec{q}_s d^2\vec{q}_{i}~ \tilde{K}(\vec{q}_{s},\vec{q}_{i}) ~ \hat{a}^{\dagger}(\vec{q}_{s})  \hat{a}^{\dagger}(\vec{q}_i)+h.c.,\label{eq:HwaveVec}
\end{equation}
where  $\vec{q}_{s(i)}$ is the transverse component of the wave vector for the signal (idler) field, $\hat{a}^{\dagger}(\vec{q})$ is the creation operator at $\vec{q}$ and $t=0$, and $g$ is the gain parameter proportional to the crystal nonlinearity $\chi^{(2)}$, the length of the nonlinear crystal $l_c$, and the maximum pump amplitude $A_p$. The kernel $K(\vec{q}_{s},\vec{q}_{i})$ determines the spatial properties of the generated light, which depends on the pump beam distribution and the phase matching condition. For a monochromatic Gaussian pump beam focused at the center of the crystal, the kernel is given by~\cite{Monken:1998et,lopez09}
\begin{align}
	\tilde{K}(\vec{q}_{s},\vec{q}_{i}) & = \exp \left({-{w_p^2 \over 4}\vert \vec{q}_s+\vec{q}_i \vert^2} \right)  \text{sinc}\left({l_c \over 4k_p} \left|\vec{q}_s - \vec{q}_i \right|^2 \right)
	, \label{eq:kernelWaveVec}
\end{align}
where $k_p$ and $w_p$ are the wavenumber and the waist of the pump beam, respectively. 
The Gaussian function originates from the pump distribution (making the anticorrelation between $\vec{q}_s$ and $\vec{q}_i$), and the sinc function is by the phase matching condition (making the correlation between $\vec{q}_s$ and $\vec{q}_i$).

The Hamiltonian $\hat{H}$ in \eq{eq:HwaveVec} can be described by the transverse position operator $\hat{a}^\dagger (\vec{x})~( = {1 \over 2\pi} \int d^2\vec{q}~ e^{-i\vec{q} \cdot \vec{x} }~\hat{a}^{\dagger}(\vec{q}))$ as well, using the inverse Fourier transform,
\begin{align}
	\hat{H} = -{i\hbar{g} \over 2} \int d^2\vec{x}_s d^2\vec{x}_{i}~ K(\vec{x}_{s},\vec{x}_{i}) ~ \hat{a}^{\dagger}(\vec{x}_{s})  \hat{a}^{\dagger}(\vec{x}_i)+h.c. \label{eq:H}
\end{align}
The associated kernel $K(\vec{x}_{s},\vec{x}_{i})$ is decomposed with HG$_{mn}$ functions ($\psi^H_{mn}(x,y)$) by approximating the sinc function in $\tilde{K}(\vec{q}_{s},\vec{q}_{i})$ to the Gaussian function ($\mathrm{sinc}(x^2) \approx \exp(-\alpha x^2)$)~\cite{Straupe:2011ju,miatto12},
\begin{align}
	K(\vec{x}_s,\vec{x}_i) = \sum_{m,n=0}^{\infty}\mu^{m+n}~ \psi^H_{mn}(\vec{x}_s) ~\psi^H_{mn}(\vec{x}_i), \label{eq:Kernel}
\end{align}
where
\begin{align}
	\psi^H_{mn}(x,y) & = {
	\exp\left(-{{ x^2+ y^2} \over w_H^2 }\right) H_m({\sqrt{2} x \over w_H })H_n({ \sqrt{2} y \over w_H }) \over {w_H \sqrt{2^{m+n-1} \pi m!n!}} } \nonumber \\
	\mu & = {1 - \sqrt{\xi} \over 1+\sqrt{\xi}} \nonumber \\
	\xi & = \alpha l_c/2z_p \nonumber \\
	w_H &= \sqrt{2} \xi^{1/4} w_p \label{eq:mu}
\end{align}
The pump beam property has been expressed in terms of the Rayleigh range $z_p= k_p w_p^2/2$, and $H_n(x)$ is the $n$-th order Hermite polynomial. The key parameters determining the kernel are a modified (by $\alpha$) focusing parameter of the pump $\xi$ and the waist size $w_H$. $\xi$ determines the eigenvalues $\mu^{m+n}$, where $\abs{\mu}^{m+n} < 1$, and $w_H$ determines the width of HG modes $\psi^H_{mn}(x,y)$. The Schmidt number $M$, quantifying the average number of modes~\cite{law04}, can be also expressed as a function of the focusing parameter $\xi$.
\beq
\centering
M = {1 \over 4} \left(\sqrt{\xi}+\sqrt{1/\xi}\right)^2. \label{eq:Schmidt}
\eeq
$M$ has the minimum value of one at $\xi=1$, and $M$ increases as $\xi$ deviates from one.

With this kernel decomposition, the Hamiltonian in \eq{eq:H} can be expressed in a decoupled form
\begin{equation}
	\hat{H} = -{i\hbar g \over 2} \sum_{mn} \mu^{m+n} \left(\hat{A}^{H}_{mn}\right)^{\dagger}\left(\hat{A}^{H}_{mn}\right)^{\dagger} + h.c., \label{eq:Hdecomp}
\end{equation}
where
\begin{equation}
	\left(\hat{A}^{H}_{mn}\right)^{\dagger} = \int d^2\vec{x} ~ \psi^H_{mn}(\vec{x}) ~ \hat{a}^{\dagger} (\vec{x})
\end{equation}
are the creation operators of the eigenmodes of the interaction Hamiltonian, which are in the $\textrm{HG}_{mn}$ modes $\psi^H_{mn}(\vec{x})$. The operators satisfy the bosonic commutation relation $[\hat{A}^{H}_{mn},\left(\hat{A}^{H}_{kl}\right)^{\dagger}] = \delta_{mn,kl}$, constructing an orthonormal HG mode basis. 
Note that the Hamiltonian is a direct sum of $\left(\hat{A}^{H}_{mn}\right)^{\dagger}\left(\hat{A}^{H}_{mn}\right)^{\dagger} + \hat{A}^{H}_{mn} \hat{A}^{H}_{mn}$ from different HG$_{mn}$ modes, which means that it generates squeezed light in each of the multiple spatial modes. The eigenvalues $\mu^{m+n}$ are associated with the relative squeezing levels in different modes, which will be discussed in the following section.

\subsection{Self-imaging OPO}\label{sec:OPO}

\begin{figure}[t]
    \centering
	\includegraphics[width=0.4\textwidth]{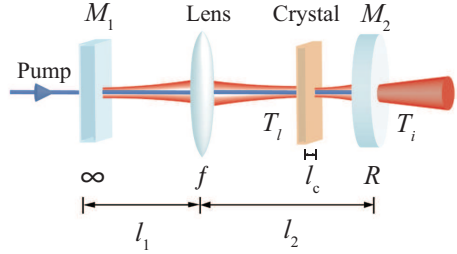}
	\caption{\label{fig:OPO}
Self-imaging OPO for generating multimode squeezed light. $M_1$ is a perfect dichroic mirror transmitting the pump and reflecting the squeezed light, and $M_2$ is a partially reflective mirror having a transmittance of $T_i$ for the squeezed light. $T_l$ is the transmittance associated with the intracavity loss. The distances between the mirrors and the lens are $l_1$ and $l_2$, as represented in \eq{eq:self-image}, where detunings $\Delta l_1$ and $\Delta l_2$ are required for stable operation of the OPO \cite{chalopin10}.}
\end{figure}

Let us construct a cavity that is compatible with the spatially multimode Hamiltonian in \eq{eq:Hdecomp}. While a typical cavity resonates on a single HG$_{00}$ mode~\cite{Vahlbruch:2016jf}, for our purpose, we require a degenerate cavity resonant on all HG$_{mn}$ modes. The cavity degeneracy is determined by the Gouy phase shift ($\theta_{G}$) accumulated by the HG$_{00}$ mode along one cavity round-trip~\cite{siegman86,Gigan:2005fn}. A confocal cavity has $\theta_{G}=\pi~ \textrm{mod}(2\pi)$, which is resonant only on half of HG$_{mn}$ modes having either an even or an odd number of $m+n$~\cite{lugiato97}. On the other hand, a self-imaging cavity exhibits $\theta_{G}=0 ~ \textrm{mod}(2\pi)$: it is a fully resonant cavity for all HG$_{mn}$ modes~\cite{Arnaud:1969gh,chalopin10}.

Figure \ref{fig:OPO} describes a self-imaging OPO, which consists of a plane mirror $M_1$, a lens of focal length $f$, a nonlinear crystal of length $l_c$, and a curved mirror $M_2$ of a radius of curvature $R$. The lengths $l_1$ and $l_2$ can be expressed as
\begin{align}
l_1 & = f+f^2/R+\Delta l_1 \nonumber \\
l_2 & = f+R+\Delta l_2.
\label{eq:self-image}
\end{align}
When $\Delta l_1 = \Delta l_2 = 0$, the OPO becomes fully degenerate for all HG$_{mn}$ modes. This ideal condition, however, leads to cavity instability~\cite{chalopin10}, and thus, small detunings ($\Delta l_1, \Delta l_2$) are required for stable operation. The Gouy phase shift with such a detuning is
\begin{equation}
	\theta_{G} = \cos^{-1} \left( 1 + 2 {\Delta l_2 \over R} \left( {\Delta l_1 \over R}  + {\Delta l_2 \over R} -  {\Delta l_1 \over R} {\Delta l_2 \over R} \right) \right),\label{eq:Gouy}
\end{equation}
where we have assumed $f=R$. As a result, when the cavity is locked for HG$_{00}$ mode, a high-order HG$_{mn}$ mode attains a phase shift of $(m+n)\theta_{G}$~\cite{siegman86}. The cavity has HG eigenmodes 
\begin{equation}
    \psi_{mn}(x,y) = {\exp \left( -{x^2+y^2 \over w_c^2}\right) H_m \left({\sqrt{2}x \over w_c}\right) H_n\left({\sqrt{2}y \over w_c}\right) \over w_c \sqrt{2^{m+n-1}\pi m! n!}}
    \label{eq:cavitymode}
\end{equation}
with the waist size of $w_c = \sqrt{R\lambda_0/2\pi}$ ($\lambda_0$: the free space wavelength) for small detunings $\Delta l_1, \Delta l_2 \ll R$, and the associated creation operators are
\begin{equation}
    \hat{A}_{mn}^\dagger = \int d^2\vec{x} ~ \psi_{mn}(\vec{x}) ~ \hat{a}^{\dagger} (\vec{x}).
\end{equation}

To match the cavity modes with the eigenmodes of the Hamiltonian in \eq{eq:Kernel}, we position the crystal at the cavity waist and set $R=2\pi w_H^2 / \lambda_0$, while a more general case of mismatch between the modes will be discussed in Section~\ref{subsection:OPOmismatch}. In this configuration, we obtain a decoupled quantum-Langevin-equation for each $\hat{A}_{mn}$~\cite{Gardiner85},
\begin{equation}
	{d \over dt}\hat{A}_{mn}(t) = {1 \over i\hbar}[\hat{A}_{mn}(t),\hat{H}] -\left(\gamma_i+\gamma_l-i\Delta_{mn}\right)\hat{A}_{mn}(t) +\sqrt{2\gamma_i}\hat{A}^i_{mn}(t) + \sqrt{2{\gamma}_l} \hat{A}_{mn}^l (t),\label{eq:Langevin}
\end{equation}
where $\hat{A}_{mn}^i$ and $\hat{A}_{mn}^l$ are the annihilation operators of the input and the intra-cavity loss modes, respectively, and the corresponding decay rates are given by $\gamma_i=T_i/2\tau$ and $\gamma_l = T_l/2\tau$ ($T_i$ and $T_l$ are shown in \fig{fig:OPO}, $\tau$: the round trip time). As we consider a cavity locked for the HG$_{00}$ mode, the cavity detuning frequency of HG$_{mn}$ mode, $\Delta_{mn}$, is given by $(m+n)\theta_{G}/\tau$. We have used the approximation $T_i, T_l, \theta_G \ll 1$.
Using \eq{eq:Hdecomp},
\begin{equation}
	{1 \over i\hbar}[\hat{A}_{mn}(t),\hat{H}] = -g \mu^{m+n} \hat{A}_{mn}^{\dagger}(t),
\end{equation}
and then, the Fourier transform of \eq{eq:Langevin} becomes
\begin{equation}
	(\gamma_i+\gamma_l-i\omega-i\Delta_{mn}) \hat{A}_{mn}(\omega) + g \mu^{m+n} \hat{A}^{\dagger}_{mn}(-\omega) = \sqrt{2\gamma_i}\hat{A}^{i}_{mn}(\omega)  +\sqrt{2\gamma_l}\hat{A}^l_{mn}(\omega) \label{eq:Fourier}
\end{equation}
at frequency $\omega$.

To investigate quantum correlations of the generated light at sidebands frequency $\omega$, we employ a vector of quadrature operators in HG$_{mn}$ modes,
\begin{equation}
\mathbf{\hat{Q}}(\omega) = [\hat{X}_{00}(\omega),\hat{X}_{01}(\omega),...,\hat{P}_{00}(\omega),\hat{P}_{01}(\omega),...]^T,
\end{equation}
where $\hat{X}_{mn}(\omega) = \hat{A}_{mn}(\omega)+\hat{A}_{mn}^{\dagger}(-\omega)$ and $\hat{P}_{mn}(\omega) = (\hat{A}_{mn}(\omega)-\hat{A}_{mn}^{\dagger}(-\omega))/i$.
Express \eq{eq:Fourier} with this vector operator as 
\begin{equation}
	\mathbf{M}(\omega)\mathbf{\hat{Q}}(\omega) = \sqrt{2\gamma_i}\mathbf{\hat{Q}}^{i}(\omega) +\sqrt{2\gamma_l} \mathbf{\hat{Q}}^{l}(\omega)
	\label{eq:Qequation}
\end{equation}
with
\begin{equation}
	\mathbf{M}(\omega) = \left[\begin{matrix}
		(\gamma_i+\gamma_l -i\omega)\mathbf{I} + \mathbf{G} &  \mathbf{D} \\
		-\mathbf{D} & (\gamma_i+\gamma_l -i\omega)\mathbf{I} - \mathbf{G}\\
		\end{matrix}
	\right],
	\label{eq:Gmatrix}
\end{equation}
where $\mathbf{I}_{kl,mn}=\delta_{kl,mn}, \mathbf{G}_{kl,mn} = g\mu^{k+l}~\delta_{kl,mn}, \mathbf{D}_{kl,mn} = \Delta_{kl}~\delta_{kl,mn}$.
Using the input-output relation at the coupler $\mathbf{\hat{Q}}^i+\mathbf{\hat{Q}}^o = \sqrt{2\gamma_i}\mathbf{\hat{Q}}$ ($\mathbf{\hat{Q}}^i$ and $\mathbf{\hat{Q}}^o$ are the quadrature vectors of input and output modes), \eq{eq:Qequation} becomes
\begin{equation}
	\mathbf{\hat{Q}}^o(\omega)=(2\gamma_i\mathbf{M}^{-1}(\omega)-\mathbf{I})\mathbf{\hat{Q}}^i(\omega)+2\sqrt{\gamma_i \gamma_l}\mathbf{M}^{-1}(\omega)\mathbf{\hat{Q}}^l(\omega).
\end{equation}
As the input is the vacuum state, we obtain the covariance matrix generated from the OPO as follows
\begin{equation}
	\mathbf{V}(\omega) = 4\gamma_i \gamma_l \mathbf{M}^{-1}(\omega)(\mathbf{M}^{-1}(-\omega))^T
	+ (2\gamma_i\mathbf{M}^{-1}(\omega)-\mathbf{I})(2\gamma_i\mathbf{M}^{-1}(-\omega)-\mathbf{I})^T. \label{eq:V}
\end{equation}
Note that the covariance matrix $\mathbf{V}(\omega)$ does not exhibit any coupling between different HG$_{mn}$ modes. We can therefore characterize it by considering each HG$_{mn}$ mode individually. The quantum state in each HG$_{mn}$ mode turns out to be a squeezed vacuum aligned with rotated quadratures $\hat{X}_{mn}^{(\Theta)}(\omega), \hat{P}_{mn}^{(\Theta)}(\omega)$:
\begin{align}
\hat{X}_{mn}^{(\Theta)}(\omega) &=\hat{X}_{mn}(\omega) \cos \Theta + \hat{P}_{mn}(\omega) \sin \Theta \nonumber \\
\hat{P}_{mn}^{(\Theta)}(\omega) &= - \hat{X}_{mn}(\omega) \sin \Theta + \hat{P}_{mn}(\omega) \cos \Theta \nonumber \\
\expec{\Delta^2 \hat{X}_{mn}^{(\Theta)}(\omega)}   &= 1 - \eta
	{ {4 \abs{\tilde{g}_{mn}}} \over
	 { 2 \abs{\tilde{g}_{mn}} + \sqrt{(\tilde{\omega}^2-\tilde{\Delta}_{mn}^2+\tilde{g}_{mn}^2+1)^2 +4  \tilde{\Delta}_{mn}^2}}} \nonumber \\
\expec{\Delta^2 \hat{P}_{mn}^{(\Theta)}(\omega)}  &= 1 - \eta
	{ {4 \abs{\tilde{g}_{mn}}} \over
	 { 2 \abs{\tilde{g}_{mn}} - \sqrt{(\tilde{\omega}^2-\tilde{\Delta}_{mn}^2+\tilde{g}_{mn}^2+1)^2 +4  \tilde{\Delta}_{mn}^2}}} \nonumber \\
\expec{ \{ \Delta \hat{X}_{mn}^{(\Theta)}(\omega), \Delta \hat{P}_{mn}^{(\Theta)}(\omega) \} }   &= 0 \nonumber \\
	\tan \Theta &=  {2 \tilde{\Delta}_{mn}	\over
	(\tilde{\omega}^2-\tilde{\Delta}_{mn}^2+ \tilde{g}_{mn}^2 +1) +\textrm{sgn}[\tilde{g}_{mn}] \sqrt{(\tilde{\omega}^2-\tilde{\Delta}_{mn}^2+ \tilde{g}_{mn}^2 +1)^2 + 4 \tilde{\Delta}_{mn}^2}}, 	\label{eq:XPmn}
\end{align}
where all the parameters are real numbers, and
\begin{align}
\eta &= \frac{\gamma_i}{\gamma_i+\gamma_l} = \frac{T_i}{T_i+T_l} \nonumber\\ 
\tilde{\omega} &= \frac{\omega} {\gamma_i+\gamma_l} \nonumber \\
\tilde{g}_{mn} &= \frac{g} {\gamma_i+\gamma_l} \mu^{m+n} \nonumber \\
\tilde{\Delta}_{mn} &= \frac{\Delta_{mn}} {\gamma_i+\gamma_l} =  \frac{2\theta_G} {T_i+T_l} (m+n). \label{eq:XPpara}
\end{align}
Note that $\expec{\Delta^2 \hat{X}_{mn}^{(\Theta)}(\omega)} \le 1$ and $\expec{\Delta^2 \hat{P}_{mn}^{(\Theta)}(\omega)} \ge 1$, and as $\tilde{\Delta}_{mn} \to 0$, the angle $\Theta$ is $0$ if $0 \le \tilde{g}_{mn}$ or $\pi/2$ if $\tilde{g}_{mn} < 0$. Without loss of generality, we will focus on the case of $0 \le \tilde{g}_{mn}$ by setting $0 \le g$ and $0 \le \mu < 1$ (equivalently, $0 < \xi \le 1$). One can further note that the intracavity loss $T_l$ makes reduction on the cavity escape efficiency $\eta$.

\begin{figure}[t]
	\includegraphics[width=120mm]{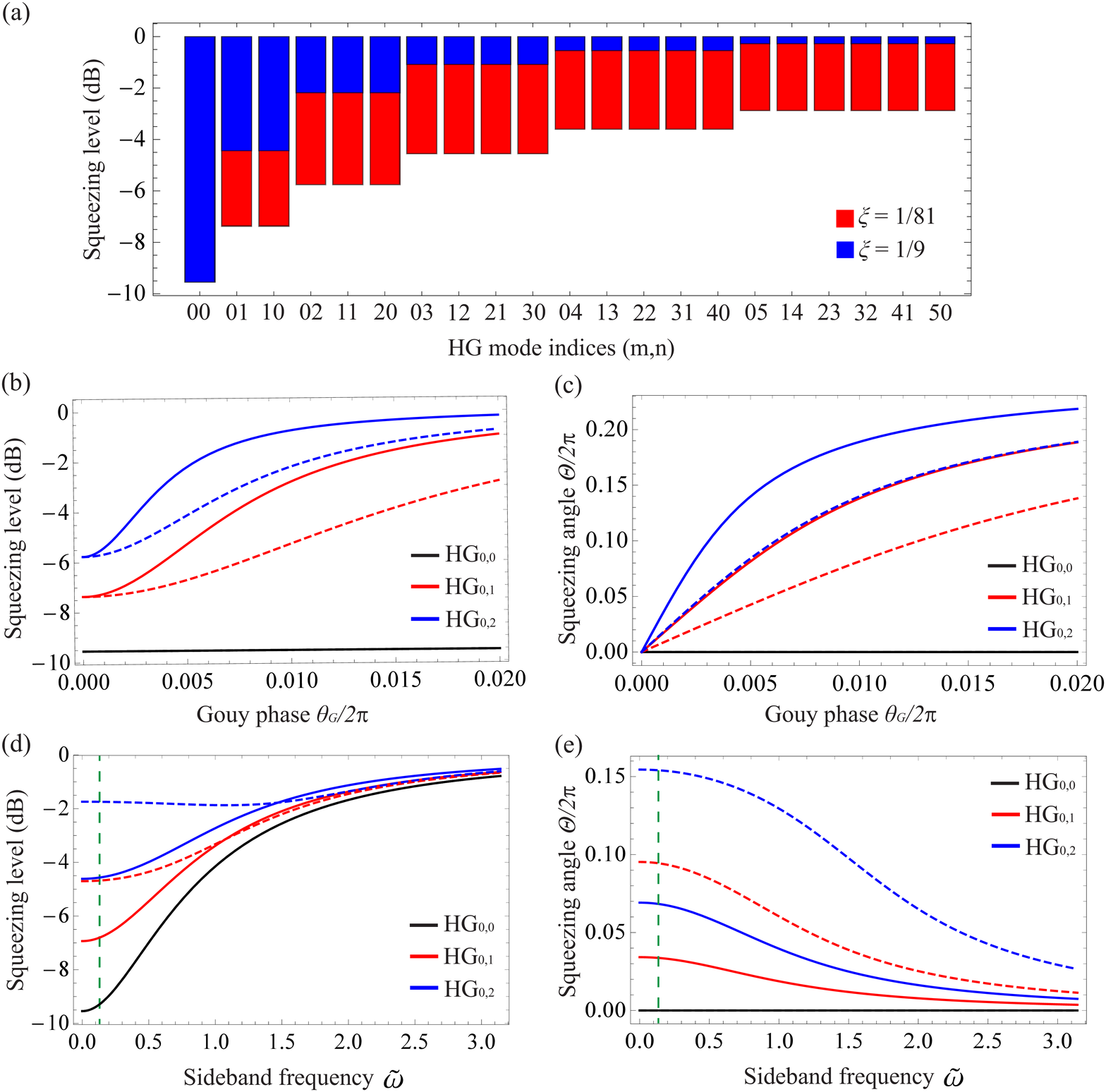}
	\centering
	\caption{\label{fig:squeezing_gouy} (a) Squeezing levels of multimode squeezed light for different focusing parameters of $\xi = 1/9$ and $\xi = 1/81$, when $T_i = 0.1, T_l = 0, \tilde{\omega} = 0, \tilde{g}_{00}=1/2$, and $\theta_G = 0$. In both cases, the squeezing levels in HG$_{00}$ are 9.5 dB. Effect of the Gouy phase shift $\theta_G$ on the squeezing levels in (b) and the squeezing angles in (c) for different HG$_{mn}$ modes and the transmittance $T_i$ of $M_2$. The solid lines are for $T_i = 0.1$, and the dashed lines are for $T_i = 0.2$. We use $\xi = 1/81$, $T_l=0$, $\tilde{\omega} =0$, and $\tilde{g}_{00} = 1/2$. Dependence of the squeezing levels in (d) and the squeezing angles in (e) on the sidebands frequency $\tilde{\omega}$. The solid and dashed lines are for $\theta_G/2\pi = 0.002$ and $\theta_G/2\pi = 0.006$, respectively. We use $\xi = 1/81$, $T_i = 0.1$, $T_l=0$, and $\tilde{g}_{00} = 1/2$. For the later analyses in Figures~(\ref{fig:mismatching}, \ref{fig:loss}, \ref{fig:enhancement}), sideband frequency of $\tilde{\omega} = \pi/25$ will be used, which is indicated as the vertical dashed lines.
	}
\end{figure}

The generated multimode light from the OPO is therefore a collection of individual squeezed vacua in multiple HG$_{mn}$ modes, whose modal structure and quantum correlations are described by \eq{eq:mu} and \eq{eq:XPmn}, respectively. In more detail,
the spectrum of quantum correlations is determined by the focusing parameter $\xi$, modifying $\tilde{g}_{mn}$ in \eq{eq:XPpara}, and the waist size $w_0$ is by $\xi$ and the pump waist $w_p$, as discussed in \eq{eq:mu}. For $\xi = 1$, making $\mu= 0$, the generated light is a single-mode squeezed vacuum:  HG$_{00}$ contains a squeezed vacuum ($\expec{\Delta^2 \hat{X}_{00}^{(\Theta)}(\omega)}< 1$), but all high-order HG$_{mn}$ modes ($m,n\ne 0$) are vacuum states ($\expec{\Delta^2 \hat{X}_{mn}^{(\Theta)}(\omega)}= 1$). On the other hand, as $\xi$ becomes smaller than one, $\mu$ becomes positive, and high-order modes also exhibit squeezing ($\expec{\Delta^2 \hat{X}_{mn}^{(\Theta)}(\omega)} < 1$). Figure \ref{fig:squeezing_gouy}(a) compares the squeezing levels $\expec{\Delta^2 \hat{X}_{mn}^{(\Theta)}(\omega)}$ for different values of $\xi$. The smaller $\xi$ exhibits higher squeezing levels than the larger one for all high orders of HG$_{mn}$, and the associated Schmidt numbers calculated from \eq{eq:Schmidt} are 8.3 ($\xi=1/9$) and 20.7 ($\xi=1/81$). The angles $\Theta$ for $\hat{X}_{mn}^{(\Theta)}(\omega)$ quadratures are all zero as expected.

Figures \ref{fig:squeezing_gouy}(b,c) show the effects of the Gouy phase shift $\theta_G$ on the generated light. The Gouy phase creates the detuning $\tilde{\Delta}_{mn}$, which affects both the squeezing level and the squeezing angle. For small detunings, the squeezing level and the squeezing angle remain similar to the ideal self-imaging case, but as $\theta_G$ increases, the squeezing level $\expec{\Delta^2 \hat{X}_{mn}^{(\Theta)}(\omega)}$ gradually degrades to zero, and the squeezing angle $\Theta$ increases to $\pi/2$. Such effects are stronger for a smaller transmittance $T_i$ and a higher HG$_{mn}$ mode, as expected from \eq{eq:XPpara}. In addition, the squeezing level and the squeezing angle depend on the sideband frequency $\tilde{\omega}$, as shown in Figs. \ref{fig:squeezing_gouy}(d,e). As HG$_{00}$ mode exhibits $\tilde{\Delta}_{mn}=0$, it behaves as a common OPO, where the squeezing level decreases while the squeezing angle remains constant as $\tilde{\omega}$ increases. On the other hand, for higher modes where $\tilde{\Delta}_{mn} \ne 0$, both of the squeezing level and the squeezing angle depend on the sideband frequency $\tilde{\omega}$. The rotated squeezing angle due to non-zero $\tilde{\Delta}_{mn}$ returns to zero as $\tilde{\omega}$ increases. The squeezing level, in most cases, gradually decreases to zero by increasing $\tilde{\omega}$, but there is a special case showing a non-monotonic behavior (the blue dashed line for HG$_{02}$ in \fig{fig:squeezing_gouy}(d)), which is because a non-zero value of $\tilde{\omega}$ makes the minimum value for $\expec{\Delta^2 \hat{X}_{mn}^{(\Theta)}(\omega)}$: such a case can take place at $\tilde{\omega} = \sqrt{\tilde{\Delta}_{mn}^2 - \tilde{g}_{mn}^2-1}$ for $\tilde{\Delta}_{mn}^2 > \tilde{g}_{mn}^2+1$, which can be derived from \eq{eq:XPmn}.

\section{Robustness on spatial mode mismatch}

Spatial mode mismatch occurs when the mode of quantum light is different from a target mode, e.g. due to beam displacement, tilting, and beam size difference. Mode mismatch is especially detrimental for couplings with single-mode elements and processes, e.g., optical cavities, optical fibers, frequency conversion, and homodyne detection. In this section, we will show that the multimode squeezed light from the self-imaging OPO is robust on various types of spatial mode mismatch. When deriving the result, we will consider mode mismatch only in the $x$-direction, but the same result can be equally obtained for the $y$-direction because of the symmetry of the multimode squeezed light described in Eqs. (\ref{eq:Kernel},\ref{eq:mu},\ref{eq:XPmn}).

\subsection{Mode-mismatch model}\label{subsection:model}

\begin{figure}[t]
	\centering
	\includegraphics[width=0.6\textwidth]{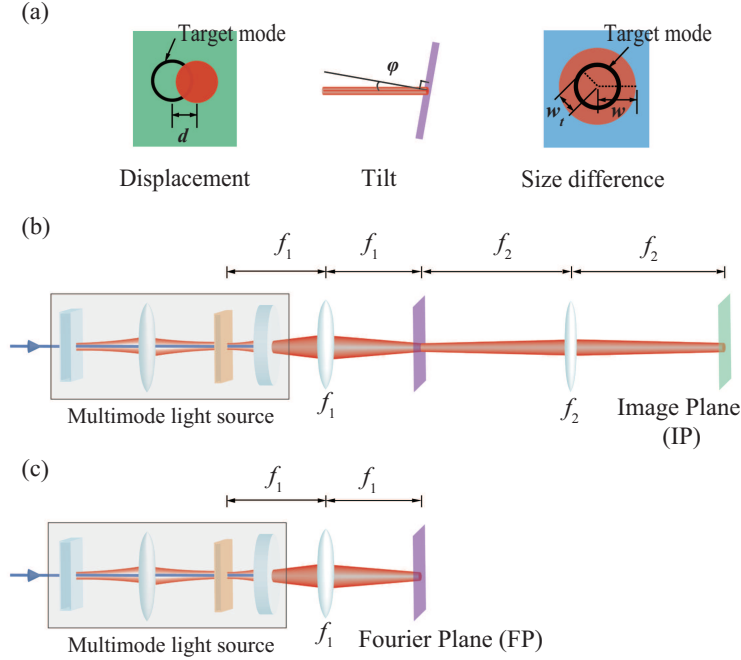}
	\caption{\label{fig:plane}
	(a) Types of spatial mode mismatch. The displacement and the tilt are drawn in the $x$-direction for clarity, but they can be generalized to arbitrary directions. (b) 4f system for transformation into the image plane (IP). Two lenses of focal lengths $f_1$ and $f_2$ are used. (c) Transformation into the Fourier plane (FP) by employing a single lens (focal length: $f_1$).}
\end{figure}

\begin{figure}[t]
\centering
\includegraphics[width=140mm]{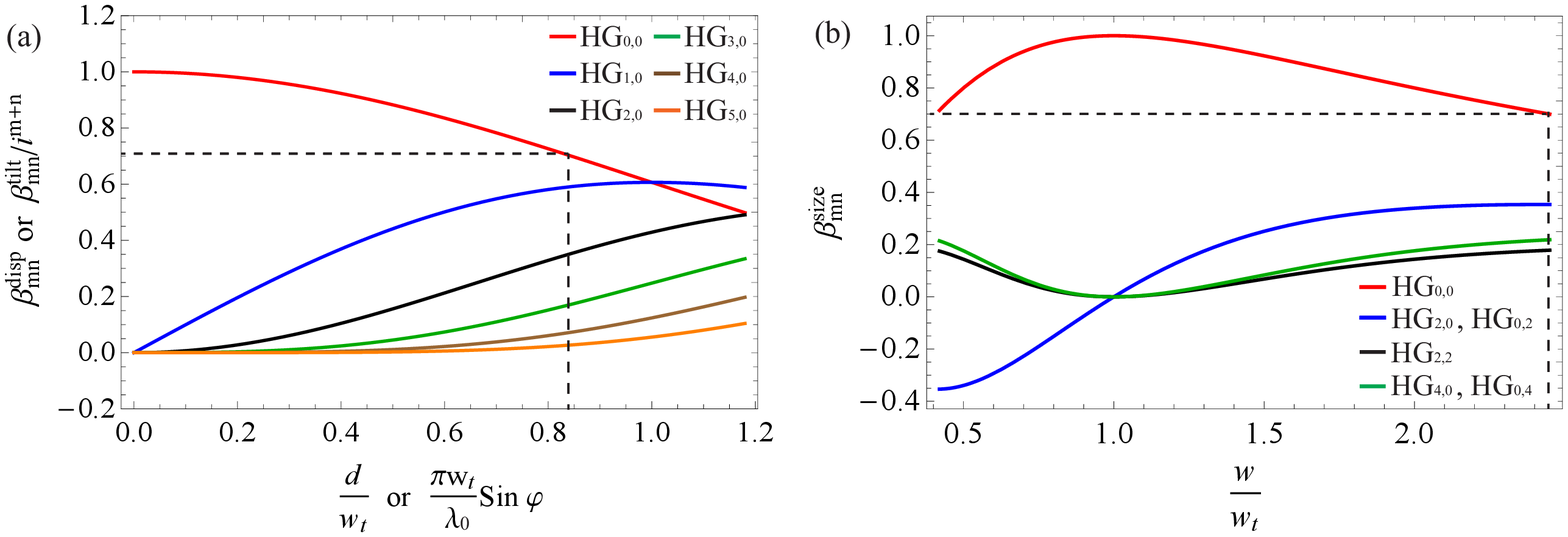}
\caption{\label{fig:mismatching_coeff} Coefficients arising from mode mismatch: displacement ($\beta_{mn}^\textrm{disp}$) and tilt ($\beta_{mn}^\textrm{tilt}$) in (a) and size difference ($\beta_{mn}^\textrm{size}$) in (b). As mode mismatch increases, the contribution of HG$_{00}$ decreases gradually while high-order HG$_{mn}$ contribute more. Dashed lines represent reduction of contribution from HG$_{00}$ by 50 \% (or $1/\sqrt{2}$ for the coefficient) due to mode mismatch, which takes place at $d/w_t = \pi w_t \sin \varphi /\lambda_0  = 0.83$ and $w/w_t = 2.45$.
}
\end{figure}

To model the spatial mode mismatch, instead of fixing a target mode and varying the modes of quantum light, we will use an equivalent way for the simplicity of mathematical description: \textit{we fix the quantum light but make deviations on the target mode.} We consider a target mode of HG$_{00}$ with the waist size of $w_t$
\begin{align}
	\phi_{00} (x,y) = \sqrt{2 \over \pi}{1 \over w_t} \exp{\left(-{x^2 +y^2 \over w_t^2}\right)},\label{eq:target}
\end{align}
and its deviations due to mode mismatches (displacement ($d$), tilt ($\varphi$), and size difference ($w$)) are
\begin{align}
	\phi^\textrm{disp} (x,y;d)&= \sqrt{2 \over \pi}{1 \over w_t} \exp{\left(-{(x-d)^2 +y^2 \over w_t^2}\right)}, \nonumber \\
	\phi^\textrm{tilt} (x,y;\varphi)&= \sqrt{2 \over \pi}{1 \over w_t} \exp \left(-{x^2+y^2 \over w_t^2} + i {2\pi \over \lambda_0} x \sin \varphi \right), \nonumber \\
	\phi^\textrm{size} (x,y;w)&= \sqrt{2 \over \pi}{1 \over w} \exp{\left(-{x^2 +y^2 \over w^2}\right)},
\end{align}
respectively. Figure~\ref{fig:plane}(a) describes the mode mismatches on a target plane. One can expand a mismatched mode $\phi^\textrm{mis}$ based on the HG modes $\phi_{mn}$ stemming from \eq{eq:target}
\begin{align}
	\phi^\textrm{mis} (x,y;p) = \sum_{mn} \beta_{mn}^\textrm{mis}(p)~ \phi_{mn} (x,y), \label{eq:phimis}
\end{align}
where $\textrm{mis} \in \{ \textrm{disp,tilt,size} \}$ and a mode-mismatching parameter $p \in \{ d, \varphi, w \}$, and
\begin{align}
\beta_{mn}^\textrm{disp} (d) &= {\delta_{n,0} \over \sqrt{m!}}  \left({d \over w_t}\right)^m  \exp \left( -{d^2 \over 2w_t^2}\right), \nonumber \\
\beta_{mn}^\textrm{tilt} (\varphi) &= i^{m+n} {\delta_{n,0} \over \sqrt{m!}}  \left({\pi w_t \sin \varphi \over \lambda_0}\right)^m  \exp \left( -{\pi^2 w_t^2 \sin ^2 \varphi \over 2 \lambda_0^2}\right), \nonumber \\
\beta_{mn}^\textrm{size} (w) &= \begin{cases} 0 & n \mbox{ or } m \mbox{: odd} \\
{\sqrt{m!n!}~\left({1\over2}\tanh(\ln {w \over w_t})\right)^{m+n \over 2} \over {m \over 2}!{n \over 2}! \cosh( \ln {w \over w_t})}& n \mbox{ and } m \mbox{: even}
\end{cases},
\end{align}
and
\begin{align}
	\phi_{mn} (x,y) ={ \exp \left({-{{x^2+y^2} \over w_t^2} }\right) H_m\left({\sqrt{2}x \over w_t}\right)H_n\left({\sqrt{2}y \over w_t}\right) \over {w_t \sqrt{2^{m+n-1} \pi m!n!}  }  }. \label{eq:targetHG}
\end{align}
Figure \ref{fig:mismatching_coeff} shows the coefficients $\beta_{mn}^\textrm{disp}$, $\beta_{mn}^\textrm{tilt} (-i)^{m+n}$, and $\beta_{mn}^\textrm{size}$, which are all real values. As $ d, \varphi, \textrm{ and } w$ deviate from the ideal mode-matching condition more, HG$_{00}$ contributes less, which is replaced by the contributions from high-order HG$_{mn}$ modes.

By defining the creation operator $(\hat{B}^\textrm{mis})^\dagger$ for mode $\phi^\textrm{mis}$ and $\hat{B}^\dagger_{mn}$ for mode $\phi_{mn}$, \eq{eq:phimis} can be expressed as
\begin{align}
(\hat{B}^\textrm{mis})^\dagger = \sum_{mn} \beta_{mn}^\textrm{mis}~ \hat{B}^\dagger_{mn}. \label{eq:Bmis}
\end{align}
The effect of mode mismatch can therefore be understood as contributions from high-order HG$_{mn}$ modes due to the emergence of non-zero coefficients $\beta_{mn}^\textrm{mis}$. More specifically, a quadrature operator for the mismatched mode is written as
\begin{equation}
 \hat{B}^\textrm{mis} + (\hat{B}^\textrm{mis})^{\dagger} = \sum_{mn} \re{\beta_{mn}^\textrm{mis}} (\hat{B}_{mn} + \hat{B}^{\dagger}_{mn}) + \im{\beta_{mn}^\textrm{mis}} (\hat{B}_{mn} - \hat{B}^{\dagger}_{mn})/i.
\label{eq:mismatchingQuadrature}
\end{equation}
When the coefficients are real ($\beta_{mn} \in \mathbb{R}$) and no correlation exists between different HG$_{mn}$ and HG$_{kl}$, i.e., $\langle \Delta (\hat{B}_{mn} + \hat{B}^{\dagger}_{mn}) \Delta (\hat{B}_{kl} + \hat{B}^{\dagger}_{kl}) \rangle = 0$, the quadrature variance in the mismatched mode is
\begin{equation}
    \langle \Delta^2 \left(\hat{B}^\textrm{mis} + (\hat{B}^\textrm{mis})^{\dagger} \right) \rangle = \sum_{mn} ~ \beta_{mn}^2 ~ \langle \Delta^2 \left( \hat{B}_{mn} + \hat{B}^{\dagger}_{mn} \right) \rangle,
\end{equation}
which is the weighted mean of the quadrature variances in the HG$_{mn}$ modes with the weighting factors of $\beta_{mn}^2$. As the mode mismatch increases, the weight for HG$_{00}$ decreases, and the noises from high-order HG$_{mn}$ come in. Since single-mode squeezed light exhibits a squeezed noise in HG$_{00}$ and the vacuum noise in HG$_{mn}$, the squeezing level quickly degrades to the vacuum noise due to mode mismatch. On the other hand, multimode squeezed light exhibits squeezed noises in high-order HG$_{mn}$ modes together. As a result, multimode light can show less degradation on the squeezing level, which, therefore, tolerates more mode mismatch than single-mode light does.

\subsection{Mode-mismatch tolerance of multimode squeezed light}
We will use the multimode squeezed light in Section \ref{section:multisqz} to investigate its robustness on mode mismatch. We first consider the ideal mode matching of the multimode light with a target mode and then, to account for mode mismatch, we will make deviations on the target mode, as discussed in section \ref{subsection:model}. Figure~\ref{fig:plane}(b,c) depicts linear optical elements through which the multimode light propagate from the OPO to a target plane. The optical elements transform the HG modes $\psi_{mn}$ in \eq{eq:cavitymode} into new modes $\mathcal{I} [\psi_{mn}]$, which can be obtained by Huygen-Fresnel's integral $\mathcal{I}$ through the associated ABCD matrix~\cite{siegman86}:
\begin{equation}
	\mathcal{I} [\psi_{mn}]
	 ={ 1 \over {w_1 \sqrt{2^{m+n-1} \pi m!n!}  }  } \left( {w_1 \over A w_c+2i B/k w_c}\right)^{m+n+1} H_m \left({\sqrt{2} x \over w_1}\right) H_n \left({\sqrt{2}y \over w_1}\right) \exp \left({i k{{x^2+y^2} \over 2 q} }\right),
\end{equation}
where $A,B,C,$ and $D$ are the matrix elements, and
\begin{align}
	q &= {-A (i k w_c^2 /2) + B \over -C (i k w_c^2 /2) + D}, \nonumber \\
	w_1^2 &= A^2 w_c^2 + (2 B/ k w_c)^2. \nonumber
\end{align}

First, let us consider the transformation into the image plane (IP) in Fig.~\ref{fig:plane}(b). By choosing focal lengths satisfying $ {f_2 / f_1} = {w_t / w_c}$, the new modes become
\beq
\mathcal{I}^{\textrm{IP}} [\psi_{mn}] =  (-1)^{m+n+1} \phi_{mn} (x,y),
\eeq
which coincides with $\phi_{mn}$ in \eq{eq:targetHG} with the additional phase factor $(-1)^{m+n+1}$.
Denoting the unitary operation for $\mathcal{I}^{\textrm{IP}}$ by $\hat{U}_\textrm{IP}$, the associated creation operators show a simple relation
\begin{align}
\hat{A}^\dagger_{mn} = (-1)^{m+n+1}~\hat{U}_\textrm{IP}^\dagger \hat{B}^\dagger_{mn} \hat{U}_\textrm{IP}.
\end{align}
Together with \eq{eq:Bmis},
\begin{align}
\hat{U}_\textrm{IP}^\dagger (\hat{B}^\textrm{mis})^\dagger \hat{U}_\textrm{IP} = \sum_{mn} \beta_{mn}^\textrm{mis}~(-1)^{m+n+1}~ \hat{A}^\dagger_{mn}.
\end{align}
We thus obtain the expression of a quadrature variance at the mismatched mode at sideband frequency $\omega$:
\begin{align}
& \expec{\Delta^2 (\hat{U}_\textrm{IP}^\dagger \hat{X}^\textrm{mis}(\omega) \hat{U}_\textrm{IP} )}  =   \mathbf{r} ^ T \mathbf{V}(\omega) \mathbf{r},
\label{eq:ipsqueezing}
\end{align}
where the sideband quadrature operator $\hat{X}^\textrm{mis}(\omega)$ is $\hat{B}^\textrm{mis}(\omega) + (\hat{B}^\textrm{mis})^\dagger(-\omega)$, the covariance matrix $\mathbf{V}(\omega)$ is given in Eqs. (\ref{eq:V},\ref{eq:XPmn}), and
\begin{align}
\mathbf{r} &= [\mathrm{Re}(\gamma_{00}),...,\mathrm{Im}(\gamma_{00}),...]^T \nonumber \\
\gamma_{mn} &=\beta_{mn}^\textrm{mis}~(-1)^{m+n+1}.
\end{align}
As $\mathbf{V}(\omega)$ contains $\hat{X}$--quadrature squeezed vacua in HG$_{mn}$ modes when $\theta_G = 0$ and $\tilde{g}_{mn} > 0$, if $\gamma_{mn} \in \mathbb{R}$, only squeezed-quadrature noises are coupled into the mismatched mode, which makes the multimode squeezed light robust on mode mismatch. In the image plane, such a condition is satisfied for mode mismatches by displacement and beam-size difference,
\begin{align}
\gamma_{mn}^\textrm{disp} &= \beta_{mn}^\textrm{disp} (-1)^{m+n+1} \in \mathbb{R} \nonumber \\
\gamma_{mn}^\textrm{size} &= \beta_{mn}^\textrm{size} (-1)^{m+n+1} \in \mathbb{R}.
\end{align}

Second, we investigate the mode mismatch in the Fourier plane (FP), described in~\fig{fig:plane}(c). The focal length of the lens is chosen as $f_1 = w_t w_c \pi / \lambda_0 $. Denoting the unitary operation for transforming into the Fourier plane by $\hat{U}_\textrm{FP}$, the associated creation operators are related as
\begin{align}
\hat{A}^\dagger_{mn} = i^{m+n+1}~ \hat{U}_\textrm{FP}^\dagger \hat{B}^\dagger_{mn} \hat{U}_\textrm{FP},
\end{align}
and thus, the variance by the sideband operator $\hat{X}^\textrm{mis}(\omega) = \hat{B}^\textrm{mis}(\omega) + (\hat{B}^\textrm{mis})^\dagger(-\omega)$ is
\begin{align}
& \expec{\Delta^2 (\hat{U}_\textrm{FP}^\dagger \hat{X}^\textrm{mis}(\omega) \hat{U}_\textrm{FP} )}  =   \mathbf{s} ^ T \mathbf{V}(\omega) \mathbf{s},
\label{eq:fpsqueezing}
\end{align}
where 
\begin{align}
\mathbf{s} &= [\mathrm{Re}(\zeta_{00}),...,\mathrm{Im}(\zeta_{00}),...]^T \nonumber \\
\zeta_{mn} &=\beta_{mn}^\textrm{mis}~(-i)^{m+n+1}.
\end{align}
Like the case of the image plane, the condition $\zeta_{mn} \in \mathbb{R}$ makes the multimode squeezed light robust on mode mismatch. In the Fourier plane, mismatches by tilt and beam-size difference with an additional $\pi/2$-phase shift satisfy the condition,
\begin{align}
\zeta_{mn}^\textrm{tilt} &= \beta_{mn}^\textrm{tilt} (-i)^{m+n+1}  e^{i \pi/2} \in \mathbb{R} \nonumber \\
\zeta_{mn}^\textrm{size} &= \beta_{mn}^\textrm{size} (-i)^{m+n+1} e^{i \pi/2} \in \mathbb{R}.
\end{align}

\begin{figure}[t]
\centering
\includegraphics[width=140mm]{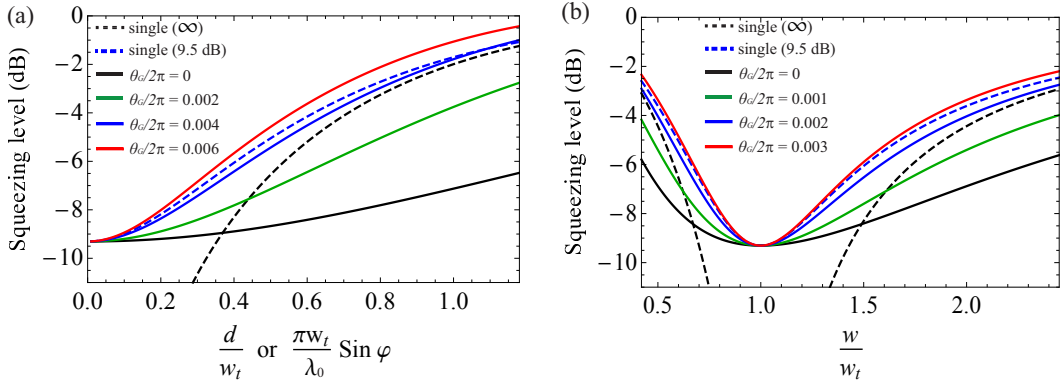}
\caption{\label{fig:mismatching} Robustness of multimode squeezed light on mode mismatch. The squeezing level coupled into a target mode is plotted by varying (a) displacement or tilt and (b) beam size. We use $T_i = 0.1$, $T_l = 0$, $\tilde{\omega} =\pi/25$, and $\tilde{g}_{00} = 1/2$. The black dashed line is for the single-mode infinitely squeezed light, and the blue dashed line is for single-mode 9.5-dB squeezed light, and the solid lines are for multimode squeezed light with $\xi = 1/81$ for different Gouy phase shifts of $\theta_G$.}
\end{figure}

Figure \ref{fig:mismatching} shows the robustness of the multimode squeezed light $\mathbf{V}(\omega)$ on mode mismatch, compared with the result of a single-mode squeezed light in HG$_{00}$. 
We first consider the multimode light by the ideal self-imaging condition ($\theta_G/2\pi=0$, black solid line), and more general cases will be discussed later. As shown in \fig{fig:mismatching}(a), when mode mismatch by displacement (in the image plane) or tilt (in the Fourier plane) occurs, the squeezing level by the single-mode light (original squeezing of 9.5 dB, blue dashed line) quickly degrades, e.g., less than 3 dB for $d/w_t > 1$ or $\pi w_t \sin \varphi / \lambda_0 > 1$. On the other hand, the multimode light with the same squeezing in HG$_{00}$ maintains the squeezing level very well by tolerating the mode mismatch, exhibiting more than 7 dB in the same condition. It is noteworthy that, at sufficiently large mode mismatch, the multimode light even outperforms single-mode light with infinite squeezing (black dashed line). Furthermore, the multimode squeezed light is robust on beam-size mismatch on both the image plane and the Fourier plane, as shown in \fig{fig:mismatching}(b). Similar to the previous case, the multimode light maintains the squeezing level very well in the influence of mode mismatch, even outperforming the single-mode infinitely squeezed light.

\subsection{Effect of loss}

\begin{figure}[t]
    \centering
	\includegraphics[width=140mm]{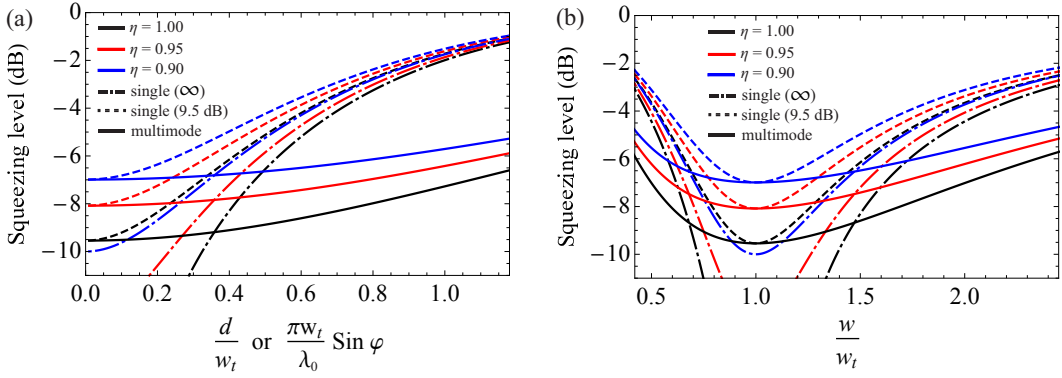}
	\caption{\label{fig:loss} Effect of loss on the squeezing level with mode mismatch. $1-\eta$ corresponds to the total optical loss (e.g. by including the detection inefficiency). For different amounts of $\eta=1$ (black), $\eta=0.95$ (red), and $\eta=0.9$ (blue), the performances of three different squeezed lights are compared (dot dashed: single-mode infinitely squeezed light, dashed: single-mode 9.5-dB squeezed light, and solid: multimode squeezed light with $\xi = 1/81$). We use the following parameters for the plots: $\tilde{\omega} =\pi/25$, $\tilde{g}_{00} = 1/2$, $\tilde{\Delta}_{mn} = 0$.}
\end{figure}

Here we investigate the effect of loss on the multimode light in terms of the mode mismatch. In Eqs. (\ref{eq:XPmn},\ref{eq:XPpara}), the escape efficiency $\eta$ accounts for the intracavity loss, but it can be generalized to incorporate the total loss in the system, $1-\eta$, e.g. propagation and detection losses. Figure~\ref{fig:loss} shows the squeezing level by mode mismatch for different amounts of losses.
$\eta=1$ corresponds to no loss in the total system ($1-\eta=0$), which is identical with the black solid lines ($\theta_G/2\pi=0$) in Figs. \ref{fig:mismatching}(a,b). As the loss increases by reducing $\eta$, the squeezing level decreases for all the three cases of infinitely squeezed single-mode light, single-mode squeezed light (9.5 dB), and the multimode light (9.5 dB in HG$_{00}$ mode). Although such losses exist, we still find that the multimode light is more robust on mode mismatch than the single-mode light (9.5 dB), and for a sufficiently large mismatch, it again outperforms the infinitely squeezed light.

\subsection{Effect of mode mismatch inside the OPO}\label{subsection:OPOmismatch}

In Section \ref{sec:OPO}, we assumed that the eigenmodes of the interaction Hamiltonian (\ref{eq:Kernel}) perfectly match with the cavity modes (\ref{eq:cavitymode}), i.e., the same waist size, $w_H = w_c$. However, mode mismatch can take place inside the OPO due to waist size difference ($w_H \ne w_c$) or the Gaussian approximation ($\mathrm{sinc}(x) \approx \exp(-\alpha x^2)$) used for the Kernel. We investigate how the mode mismatch inside the OPO affects the robustness of multimode light on mode mismatch to a target mode.

\begin{figure}[tb]
    \centering
	\includegraphics[width=140mm]{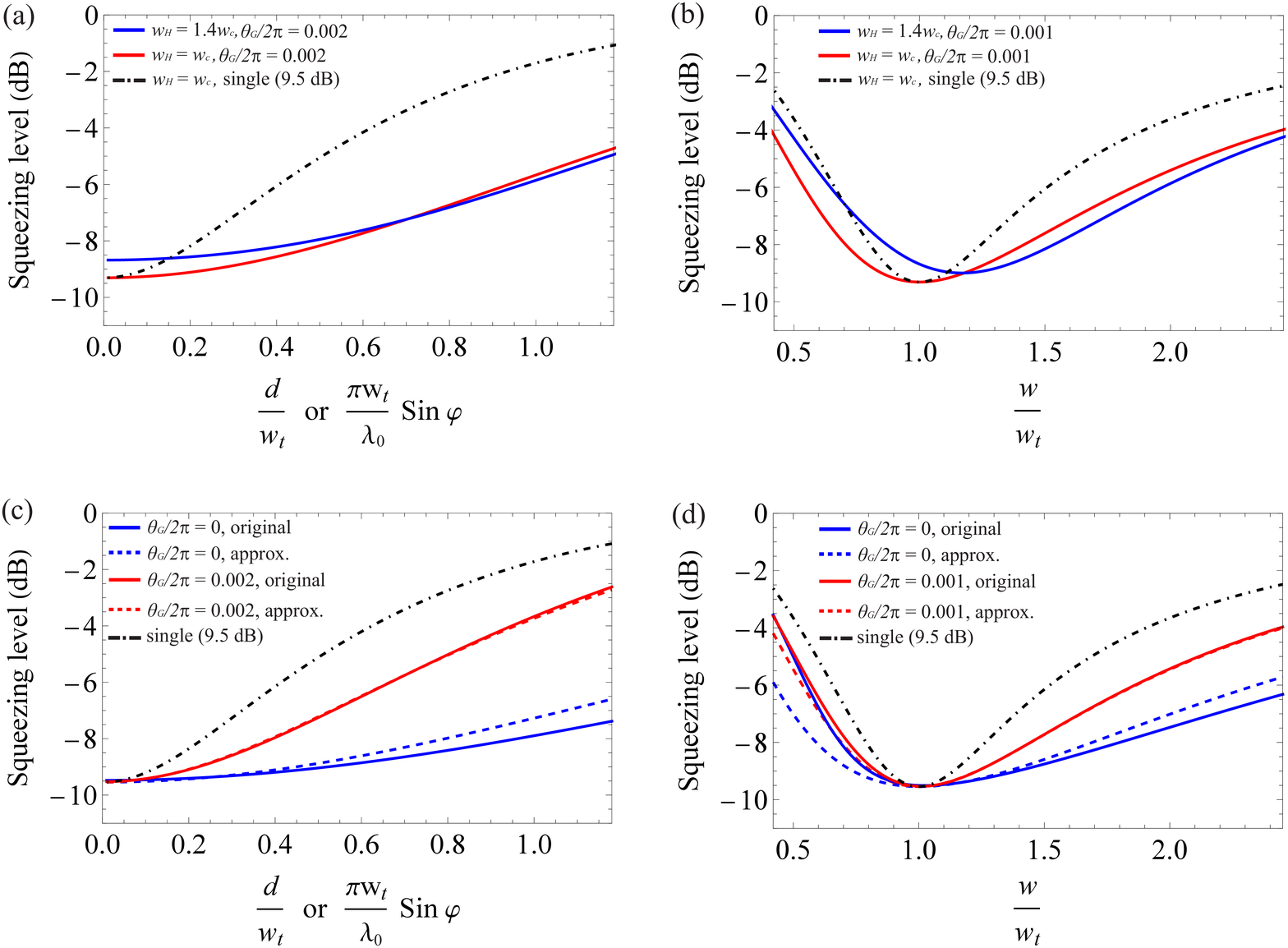}
	\caption{\label{fig:cavitymismatching} (a,b) Effect of mode mismatch due to difference in the cavity waist ($w_c$) and the interaction Hamiltonian waist ($w_H$). The squeezing level due to a large difference in waists ($w_H = 1.4 w_c$, solid blue) is compared with the case of no difference in waists ($w_H = w_c$, solid red) and with the single-mode 9.5-dB squeezed light (black dot dashed). (c,d) Effect of the Gaussian approximation (\ref{eq:Kernel}) for the original interaction Hamiltonian (\ref{eq:kernelWaveVec},\ref{eq:newKer}). The solid lines are for the original Hamiltonian and the dashed lines are for the Gaussian approximation, which are plotted without a Gouy phase (blue) and with a Gouy phase (red). Black dot-dashed line is for the single-mode 9.5-dB squeezed light. We use the following parameters for the plots: $\xi = 1/81, T_i = 0.1$, $T_l = 0$, $\tilde{\omega} =\pi/25$, $\tilde{g}_{00} = 1/2$, $\tilde{\Delta}_{mn} = 0$.		}
\end{figure}

At first, we consider the waist size difference ($w_H \ne w_c$) while keeping the Gaussian approximation. To deal with the size difference, we employ a change of basis from the eigenmodes of the interaction Hamiltonian to the cavity modes

\begin{equation}
    \psi_{mn}(x,y) = \sum_{m',n'} \mathbf{U}_{mn,m'n'}(w_c,w_H) \psi^H_{m',n'}(x,y)
\end{equation}
where the basis change matrix $\mathbf{U}(w_c,w_H)$ is given as~\cite{Kim89}
\begin{align}
\mathbf{U}_{mn,m'n'}(w_c,w_H) & = \begin{cases}
{\sqrt{m! m'! n! n'!} \over \cosh^{m+n+1} \left(\ln {w_H \over w_c}\right)} \left({\tanh\left(\ln {w_H \over w_c}\right)\over 2}\right)^{m'+n'-m-n\over 2} f\left(\ln {w_H \over w_c},m,m'\right)f\left(\ln {w_H \over w_c},n,n'\right) & \vert m-m'\vert \; \text{and} \; \vert n-n'\vert \text{: even} \\
0 & \text{else}
\end{cases} \\
f(r,m,m') &= \sum_{ {m-m'\over 2} \le k \le {m\over 2} } ~ {(-1)^k \left({\sinh r \over 2}\right)^{2k} \over k!(m-2k)!\left(k+{m-m' \over 2}\right)}.  
\end{align}
By describing the interaction Hamiltonian in the cavity mode basis, one obtains a modified gain matrix $\mathbf{G'}$
\begin{equation}
    \mathbf{G'} = \mathbf{U}(w_c,w_H)~\mathbf{G}~\mathbf{U}^{\dagger}(w_c,w_H),
\end{equation}
where $\mathbf{G}$ is the original gain matrix in Eq.~(\ref{eq:Gmatrix}). Differently from $\mathbf{G}$, $\mathbf{G'}$ is a non-diagonal matrix in general. One can use $\mathbf{G'}$ instead of $\mathbf{G}$ for calculating the covariance matrix (\ref{eq:V}) and the squeezing levels in target modes (\ref{eq:ipsqueezing},\ref{eq:fpsqueezing}).

 Figure~\ref{fig:cavitymismatching} (a,b) shows that, even with a large difference in waist sizes ($w_H = 1.4 w_c$), the light from the self-imaging OPO still exhibits robustness on mode mismatch:  displacement or tilt in Fig.~\ref{fig:cavitymismatching}(a) and beam size in Fig.~\ref{fig:cavitymismatching}(b). This robustness is due to the multimode nature of  the interaction Hamiltonian: although the waist size of the interaction Hamiltonian varies, the interaction Hamiltonian can still provide a multimode gain ($\mathbf{G'}$) in the multiple cavity modes, which in turn generates multimode squeezed light required for robustness on mode mismatch.

Second, we consider the interaction Hamiltonian without using the Gaussian approximation. Let us rewrite the associated kernel (\ref{eq:kernelWaveVec}) using $w_p$, $\xi$, and $\alpha$:
\begin{equation}
    \tilde{K}(\vec{q}_s,\vec{q}_i) = \exp \left(-{w_p^2\over 4} \vert \vec{q}_s+\vec{q}_i\vert^2\right) \mathrm{sinc} \left({w_p^2 \over 4} {\xi \over \alpha} \vert \vec{q}_s-\vec{q}_i\vert^2\right).\label{eq:newKer}
\end{equation}
We decompose the kernel numerically since analytical expression is unknown due to the inclusion of the sinc function~\cite{Straupe:2011ju}. The Schmidt number solely depends on $\xi / \alpha$ because $w_p$ just acts as the scaling factors of $\vec{q}_s$ and $\vec{q}_i$. To compare the properties of the original Hamiltonian (\ref{eq:kernelWaveVec},\ref{eq:newKer}) and those by the approximated one (\ref{eq:Kernel}), we find, for a given $\xi$, the coefficient $\alpha$ that gives the same Schmidt number as the Gaussian approximation (\ref{eq:Schmidt}); this way of choosing $\alpha$ is justified because the robustness on mode mismatch originates from the occupation of squeezed light in multiple modes, depending highly on the Schmidt number. For $\xi=1/81$, the corresponding $\alpha$ is 0.46. In addition, $w_p$ is determined by maximizing the overlap between the first eigenmode of Eq. (\ref{eq:newKer}) and the HG$_{00}$ cavity mode. A modified gain matrix $\mathbf{G'}$ is then obtained by
\begin{equation}
    \mathbf{G'}_{mn,m'n'} = g\int d^2\vec{x}_s d^2\vec{x}_i ~ K'(\vec{x}_s,\vec{x}_i)\psi_{mn}(\vec{x}_s)\psi_{m'n'}(\vec{x}_i),
    \label{eq:sincG}
\end{equation}
where $K'(\vec{x}_s,\vec{x}_i)$ is the inverse Fourier transform of Eq. (\ref{eq:newKer}), and $\psi_{mn}(x,y)$ are the cavity modes defined in Eq. (\ref{eq:cavitymode}). $\mathbf{G'}$, being a non-diagonal matrix, is used instead of $\mathbf{G}$ to find the covariance matrix (\ref{eq:V}) and the squeezing levels in target modes (\ref{eq:ipsqueezing},\ref{eq:fpsqueezing}).

In Fig.~\ref{fig:cavitymismatching}(c,d), we compare the robustness of mode mismatch by the original Hamiltonian and by the approximated Hamiltonian. It is evident that both cases exhibit robustness on mode mismatch by outperforming the single-mode squeezed light. In Fig.~\ref{fig:cavitymismatching}(c) with $\theta_G/2\pi = 0$, one can find that the original Hamiltonian shows a slightly better performance than the approximated one. It is because, while the Schmidt numbers are the same, the eigenvalue spectrum of the original Hamiltonian is distributed more toward lower-order eigenmodes than that of the approximated Hamiltonian, which is advantageous for a small amount of mode mismatch. At $\theta_G/2\pi = 0.002$, their difference becomes negligible because of the squeezing angle rotation in high-order modes. In Fig.~\ref{fig:cavitymismatching}(d) with $\theta_G/2\pi = 0$, the original Hamiltonian shows better squeezing level for $w/w_t > 1$ but worse for $w/w_t < 1$ when compared with the approximated Hamiltonian: the asymmetry comes from the negative correlations between even-order HG$_{mn}$ modes for the original Hamiltonian (e.g., $\mathbf{G'}_{00,04}$, $\mathbf{G'}_{00,22}$ < 0 in Eq.~(\ref{eq:sincG})). At $\theta_G/2\pi = 0.001$, the difference in the performances of the two Hamiltonians is negligible as in the case of Fig.~\ref{fig:cavitymismatching}(c).

\subsection{Effect of Gouy phase shift}

Until now, we have explored the robustness on mode mismatch in the ideal self-imaging condition. For stable operation of the OPO, however, a small detuning by the Gouy phase shift is necessary~\cite{chalopin10}. The detuning degrades the squeezing level and rotates the squeezing angle for high-order HG modes, as discussed in~\fig{fig:squeezing_gouy}. Here we investigate whether the robustness on mode mismatch can still be sustained with detunings from the ideal condition.

Figure~\ref{fig:mismatching} compares the squeezing levels coupled in a target mode with non-zero Gouy phase shifts, $\theta_G / 2\pi = 0.002, 0.004$, and $0.006$ for the displace/tilt mismatching and $\theta_G / 2\pi = 0.001, 0.002$, and $0.003$ for the size mismatching. As expected, the squeezing level becomes degraded as more Gouy phase shift is introduced. For $\theta_G / 2\pi = 0.002$ in (a) and $\theta_G / 2\pi = 0.001$ in (b), the generated multimode light can still beat the performance of the infinitely squeezed single-mode light for a sufficiently large mismatch. The cases of $\theta_G / 2\pi = 0.004$ in (a) and $\theta_G / 2\pi = 0.002$ in (b) exhibit a squeezing level worse than the infinitely squeezed light but better than 9.5 dB squeezed single-mode light. However, we find no advantage in using the multimode light with $\theta_G / 2\pi = 0.006$ in (a) and $\theta_G / 2\pi = 0.003$ in (b), being worse than the 9.5 dB single-mode light; in this regime, due to the rotations of the squeezing angles of high-order HG modes in~\fig{fig:squeezing_gouy}(c), $\hat{X}_{mn}(\omega)$ quadratures exhibit larger noises than the vacuum noise. To take advantage of using multimode light, keeping a small-enough Gouy phase shift is required.

\begin{figure}[tb]
    \centering
	\includegraphics[width=170mm]{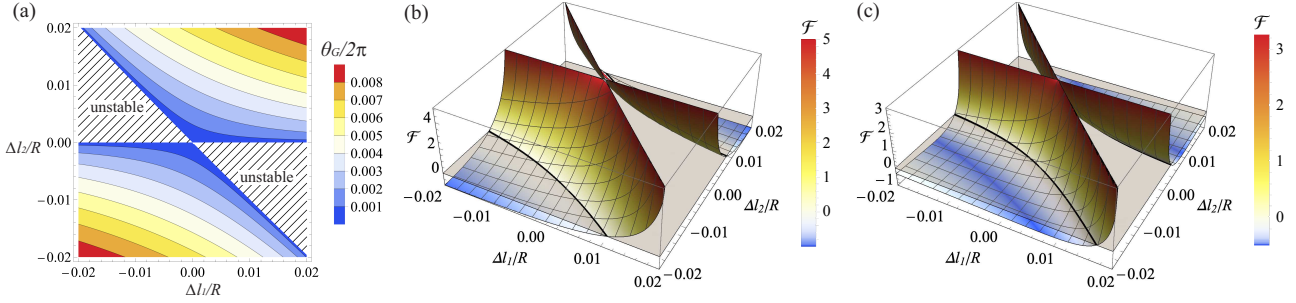}
	\caption{\label{fig:enhancement} (a) Gouy phase shift ($\theta_G$) of the OPO with detunings $\Delta l_1$ and $\Delta l_2$. Hatched areas show the unstable regions. (b,c) Enhancement factor $\mathcal{F}$ of using multimode squeezed light, defined in~\eq{eq:F}. When mode mismatch is 50 \% by (b) displacement or tilt and by (c) size difference, we compare the squeezing levels in a target mode by single-mode squeezed light (9.5 dB) and multimode squeezed light (9.5 dB at HG$_{00}$ and $\xi = 1/81$).  $\Delta l_1$ and $\Delta l_2$ are the detunings applied for cavity stability. $\mathcal{F}>0$ indicates that the multimode squeezed light is better than the single-mode light. The area where the graph is not drawn corresponds to an unstable region of the OPO. For the plot, we use parameters of $T_i = 0.1$, $T_l = 0$, $\tilde{\omega} =\pi/25$, and $\tilde{g}_{00} = 1/2$.} 
\end{figure}

We, therefore, investigate how small Gouy phase shift is required to exhibit advantages of using multimode squeezed light.
For the quantification, we define an enhancement factor $\mathcal{F}$ in decibels,
\begin{align}
\mathcal{F} = -10 \log _{10} \left[ { \Delta^2 \hat{X}^\textrm{(mul)}  \over \Delta^2 \hat{X}^\textrm{(sin)} } \right],\label{eq:F}
\end{align}
where $\Delta^2 \hat{X}^\textrm{(sin)}$ is the squeezing level by single-mode squeezed light, and $\Delta^2 \hat{X}^\textrm{(mul)}$ is the squeezing level by multimode squeezed light (having the same initial squeezing level in HG$_{00}$ as the single-mode light). A positive value of $\mathcal{F}$ indicates that the multimode squeezed light is more robust to mode mismatch than the single-mode squeezed light. As the Gouy phase shift is determined by the detuning ratios $\Delta l_1 /R$ and $\Delta l_2 /R$, as given in \eq{eq:Gouy}, we calculate the enhancement factor as varying the detunings. Figure~\ref{fig:enhancement} shows the enhancement factor of using the multimode squeezed light when the mode overlap by mismatch is 50\% (corresponding to the dashed lines at $d/w_t = \pi w_t \sin \varphi /\lambda_0 =0.83$ and $w/w_t=2.45$ in \fig{fig:mismatching_coeff}). A broad range of $\Delta l_1 /R$ and $\Delta l_2 /R$ exhibits enhancements compared with the single-mode case. Comparing (a) and (b) in \fig{fig:enhancement}, mode mismatch by size difference requires more stringent conditions for the enhancement; this is because size difference involves higher-orders of HG modes than those by displacement and tilt, as shown in~\fig{fig:mismatching_coeff}, and the squeezed lights in higher-order modes are more susceptible to the Gouy phase shift as shown in~\fig{fig:squeezing_gouy}(b,c). However, these stringent conditions are still achievable using only off-the-shelf positioning devices: a typical OPO employs a curved mirror with the radius of the curvature in the order of $R=$100 mm, and position controllability in the order of 100 $\mu m$ (e.g. by using a linear stage) can readily achieve very small values of $\Delta l_1 /R, \Delta l_2 /R = 0.001$.

\section{Conclusion}
In this paper, we have shown that multimode squeezed light generated from a self-imaging OPO is robust on spatial mode mismatch. First, we found an analytic form of the quantum properties of the multimode light at sidebands frequency by taking into account the Gouy phase shift (required for OPO stability) and the intracavity loss. By decomposing mode mismatches of displacement, tilt, and size difference into a HG$_{mn}$ mode basis, we found that the mode mismatches induce contributions from high-order HG$_{mn}$ modes, which makes the multimode squeezed light robust on mode mismatch. We showed that the multimode light from the self-imaging OPO with a small Gouy phase shift can even outperform the single-mode infinitely squeezed light, in terms of displacement and size difference in the image plane and of tilt and size difference in the Fourier plane. Such robustness on the multiple cases of mode mismatch is made possible because of the fully degenerate nature of the self-imaging OPO, which cannot be accomplished by using a confocal OPO~\cite{lugiato97} or two single-mode OPOs~\cite{steinlechner18}. Our work of mitigating the mode mismatching loss will have broad applications to quantum technologies based on squeezed light, e.g., quantum-enhanced gravitational-wave detection~\cite{Acernese:2019hg,Tse:2019jp}, deterministic quantum teleportation~\cite{Liu:2020gh}, measurement-based quantum computing~\cite{Asavanant:2019iw,Larsen:2019gg,Ra:2020gg,Pfister:2019ck}, and Gaussian boson sampling~\cite{Arrazola:2021fx,Zhong:2020cu}.


\begin{thebibliography}{8}

\bibitem{Asavanant:2019iw}
Asavanant, W. \emph{et~al.}, {Generation of time-domain-multiplexed
  two-dimensional cluster state}, \emph{Science (New York, NY)} \textbf{366},
  373--376 (2019).

\bibitem{Larsen:2019gg}
Larsen, M.~V., Guo, X., Breum, C.~R., Neergaard-Nielsen, J.~S. and Andersen,
  U.~L., {Deterministic generation of a two-dimensional cluster state},
  \emph{Science (New York, NY)} \textbf{366}, 369--372 (2019).

\bibitem{Ra:2020gg}
Ra, Y.-S. \emph{et~al.}, {Non-Gaussian quantum states of a multimode light
  field}, \emph{Nat. Phys.} \textbf{16}, 144--147 (2020).

\bibitem{Pfister:2019ck}
Pfister, O., {Continuous-variable quantum computing in the quantum optical frequency comb}, \emph{J. Phys. B: At. Mol. Opt. Phys.} \textbf{53},
  012001 (2020).
  
\bibitem{Zhong:2020cu}
Zhong, H.-S. \emph{et~al.}, {Quantum computational advantage using photons},
  \emph{Science (New York, NY)} \textbf{370}, 1460--1463 (2020).

\bibitem{Arrazola:2021fx}
Arrazola, J.~M. \emph{et~al.}, {Quantum circuits with many photons on a
  programmable nanophotonic chip}, \emph{Nature} \textbf{591}, 54--60 (2021).

\bibitem{Liu:2020gh}
Liu, S., Lou, Y. and Jing, J., {Orbital angular momentum multiplexed
  deterministic all-optical quantum teleportation}, \emph{Nat. Commun.}
  \textbf{11}, 3875 (2020).

\bibitem{giovannetti04}
Giovannetti, V., Lloyd, S. and Maccone, L., Quantum-enhanced measurements:
  beating the standard quantum limit, \emph{Science} \textbf{306}, 1330--1336
  (2004).

\bibitem{taylor13}
Taylor, M.~A. \emph{et~al.}, Biological measurement beyond the quantum limit,
  \emph{Nat. Photonics} \textbf{7}, 229--233 (2013).

\bibitem{Guo:2019el}
Guo, X. \emph{et~al.}, {Distributed quantum sensing in a continuous-variable
  entangled network}, \emph{Nat. Phys.} \textbf{16}, 281--284 (2020).

\bibitem{Tse:2019jp}
Tse, M. \emph{et~al.}, {Quantum-Enhanced Advanced LIGO Detectors in the Era of
  Gravitational-Wave Astronomy}, \emph{Phys. Rev. Lett.} \textbf{123}, 231107
  (2019).

\bibitem{Acernese:2019hg}
Acernese, F. \emph{et~al.}, {Increasing the Astrophysical Reach of the Advanced
  Virgo Detector via the Application of Squeezed Vacuum States of Light},
  \emph{Phys. Rev. Lett.} \textbf{123}, 231108 (2019).

\bibitem{Vahlbruch:2016jf}
Vahlbruch, H., Mehmet, M., Danzmann, K. and Schnabel, R., {Detection of 15~dB
  Squeezed States of Light and their Application for the Absolute Calibration
  of Photoelectric Quantum Efficiency}, \emph{Phys. Rev. Lett.} \textbf{117},
  110801 (2016).

\bibitem{Oelker:2014kv}
Oelker, E., Barsotti, L., Dwyer, S., Sigg, D. and Mavalvala, N., {Squeezed
  light for advanced gravitational wave detectors and beyond}, \emph{Opt.
  Express} \textbf{22}, 21106--21121 (2014).

\bibitem{Toyra:2017bg}
T{\"o}yr{\"a}, D. \emph{et~al.}, {Multi-spatial-mode effects in
  squeezed-light-enhanced interferometric gravitational wave detectors},
  \emph{Phys. Rev. D} \textbf{96}, 022006 (2017).

\bibitem{steinlechner18}
Steinlechner, S. \emph{et~al.}, {Mitigating Mode-Matching Loss in Nonclassical
  Laser Interferometry}, \emph{Phys. Rev. Lett.} \textbf{121}, 263602 (2018).

\bibitem{Wittel:2015wy}
Wittel, H., \emph{Active and passive reduction of high order modes in the gravitational wave detector GEO 600}(PhD Thesis, 2015). Preprint at https://arxiv.org/abs/.

\bibitem{lugiato97}
Lugiato, L.~A. and Grangier, P., {Improving quantum-noise reduction with
  spatially multimode squeezed light}, \emph{J Opt Soc Am B} \textbf{14},
  225--231 (1997).

\bibitem{embrey15}
Embrey, C.~S., Turnbull, M.~T., Petrov, P.~G. and Boyer, V., {Observation of
  Localized Multi-Spatial-Mode Quadrature Squeezing}, \emph{Phys. Rev. X}
  \textbf{5}, 1097 (2015).

\bibitem{Hsu:2004kz}
Hsu, M. T.~L., Delaubert, V., Lam, P.~K. and Bowen, W.~P., {Optimal optical
  measurement of small displacements}, \emph{J. Opt. B: Quantum Semiclass.
  Opt.} \textbf{6}, 495--501 (2004).

\bibitem{Arnaud:1969gh}
Arnaud, J.~A., {Degenerate optical cavities}, \emph{Applied Optics} \textbf{8},
  189--196 (1969).

\bibitem{chalopin10}
Chalopin, B., Chiummo, A., Fabre, C., Ma{\^\i}tre, A. and Treps, N., Frequency
  doubling of low power images using a self-imaging cavity, \emph{Opt. Express}
  \textbf{18}, 8033--8042 (2010).

\bibitem{lopez09}
Lopez, L. \emph{et~al.}, {Multimode quantum properties of a self-imaging
  optical parametric oscillator: Squeezed vacuum and
  Einstein-Podolsky-Rosen-beams generation}, \emph{Phys. Rev. A} \textbf{80},
  043816 (2009).

\bibitem{Chalopin:2010co}
Chalopin, B., Scazza, F., Fabre, C. and Treps, N., {Multimode nonclassical
  light generation through the optical-parametric-oscillator threshold},
  \emph{Phys. Rev. A} \textbf{81}, 061804 (2010).

\bibitem{chalopin11}
Chalopin, B., Scazza, F., Fabre, C. and Treps, N., Direct generation of a
  multi-transverse mode non-classical state of light, \emph{Opt. Express}
  \textbf{19}, 4405--4410 (2011).

\bibitem{Caspani:2010jq}
Caspani, L., Brambilla, E. and Gatti, A., {Tailoring the spatiotemporal
  structure of biphoton entanglement in type-I parametric down-conversion},
  \emph{Phys. Rev. A} \textbf{81}, 033808 (2010).

\bibitem{Monken:1998et}
Monken, C.~H., Ribeiro, P. H.~S. and P{\'a}dua, S., {Transfer of angular
  spectrum and image formation in spontaneous parametric down-conversion},
  \emph{Phys. Rev. A} \textbf{57}, 3123--3126 (1998).

\bibitem{Straupe:2011ju}
Straupe, S.~S., Ivanov, D.~P., Kalinkin, A.~A., Bobrov, I.~B. and Kulik, S.~P.,
  {Angular Schmidt modes in spontaneous parametric down-conversion},
  \emph{Phys. Rev. A} \textbf{83}, 060302 (2011).

\bibitem{miatto12}
Miatto, F.~M., Brougham, T. and Yao, A.~M., {Spatial Schmidt modes generated in parametric down-conversion}, \emph{Eur. Phys. J. D} \textbf{66},
  183 (2012).

\bibitem{law04}
Law, C.~K. and Eberly, J.~H., Analysis and interpretation of high transverse
  entanglement in optical parametric down conversion, \emph{Phys. Rev. Lett.}
  \textbf{92}, 127903 (2004).

\bibitem{siegman86}
Siegman, A.~E., \emph{Lasers} (University Science Books, 1986). Preprint at https://arxiv.org/abs/.

\bibitem{Gigan:2005fn}
Gigan, S., Lopez, L., Treps, N., Ma{\^\i}tre, A. and Fabre, C., {Image
  transmission through a stable paraxial cavity}, \emph{Phys. Rev. A}
  \textbf{72}, 023804 (2005).

\bibitem{Gardiner85}
Gardiner, C.~W. and Collett, M.~J., {Input and output in damped quantum systems: Quantum stochastic differential equations and the master equation}, \emph{Phys. Rev. A} \textbf{31}, 3761--3774
  (1985).

\bibitem{Kim89}
Kim, M. S., De Oliveira, F. A. M., and Knight, P. L., {Properties of squeezed number states and squeezed thermal states},
  \emph{Phys. Rev. A} \textbf{40}, 2494 (1989).

\end{thebibliography}
\end{document}